\documentclass[aps, prb, twocolumn, preprintnumbers, amsmath, amssymb]{revtex4}
\usepackage[T2A]{fontenc}
\usepackage[utf8]{inputenc}
\usepackage[english, russian]{babel}

\usepackage{graphicx}
\usepackage{xcolor}

\begin{document}
\title{Melting and polymorphic transitions in liquid}

\author{S.M. Stishov}
\email{stishovsm@lebedev.ru}
\affiliation{P.N. Lebedev Physical Institute, Leninskii pr. 53, 119991 Moscow, Russia}

\begin{abstract}
Review of the author's data, partly still unpublished, on studying of liquid
tellurium and cesium are given. No proofs indicating phase transitions in
liquids were found. New developments in studying the liquid-liquid phase transition are briefly described.  
Some relevant ethical problems are exposed in the bibliography section. 
\end{abstract}

\maketitle

\begin{figure}[htb]
\includegraphics[width=80mm]{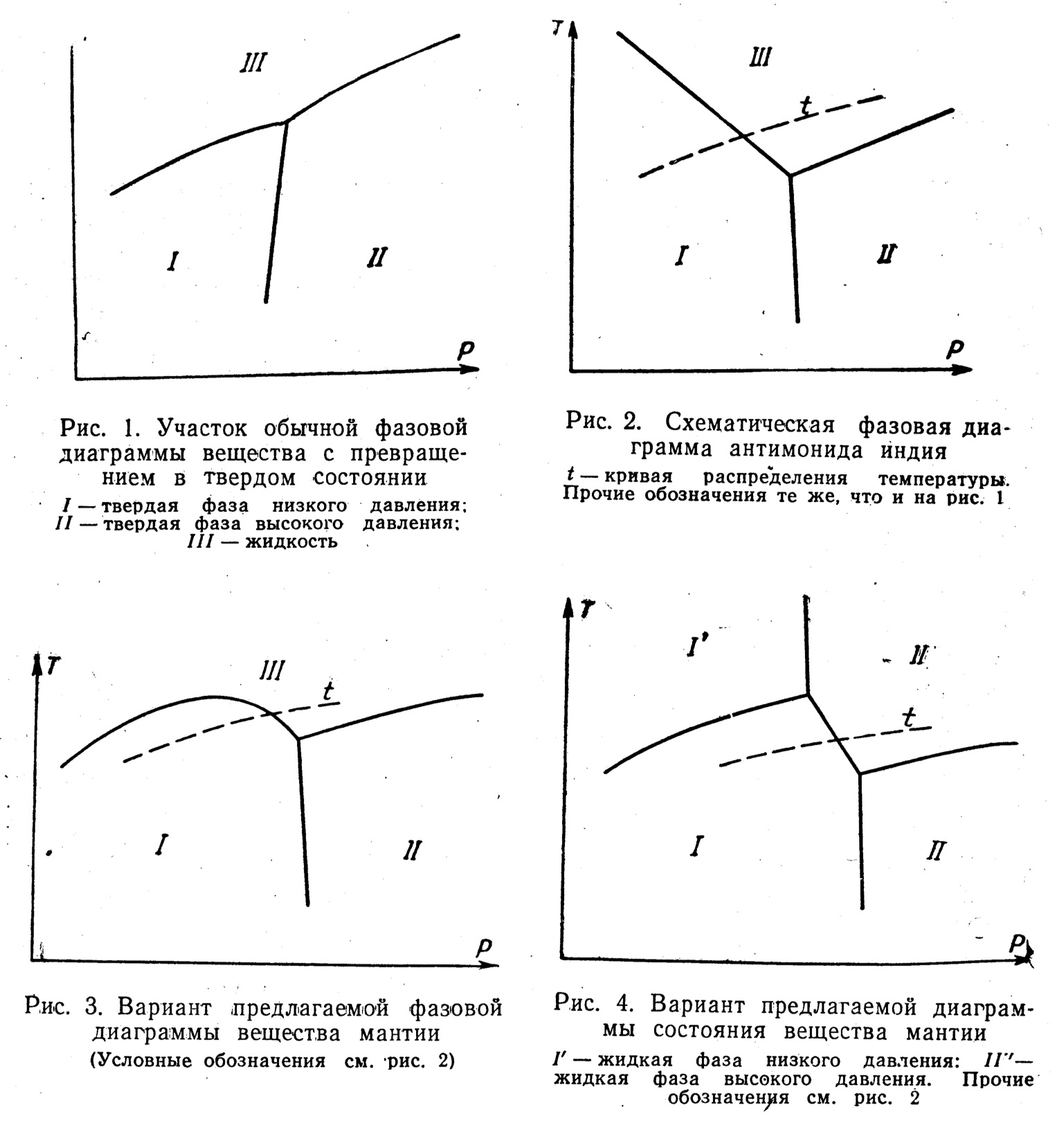}
\caption{\label{fig1} Variants of phase diagrams for one-component systems. Left lower corner: 
A phase diagram with a melting curve maxima, Right lower corner: a phase diagram with a first order
 phase transition in liquid. Reproduced from the paper, published in Russian~\cite{stish}.}
\end{figure}

My interest to the problem of phase transformations in liquid arose many
years ago, when I was deeply involved in studying the Earth structure. That
time I was fascinated by the Ramsey hypothesis claiming that the seismic
velocities and density discontinuities at the mantle-core boundary were a
result of the phase transition of the mantle material to the metallic state~ \cite{Ram}.
However, having taken into consideration that low mantle was a solid and
upper core was a liquid I suggested some sort of nontrivial phase diagrams,
which might satisfy necessary conditions~\cite{stish} (Fig.~\ref{fig1}). The point was how to
realize a coexistence line for low-pressure solid and high-pressure liquid
phases. As seen from Fig.~\ref{fig1}, a phase diagram with melting maximum and a phase
diagram with a first order phase transition in liquid could meet the
corresponding restrictions~\cite{note}. Note an existence of melting maxima may mean a
continuous transformation in liquid.

So the first thing what I had to do was to try to discover phase diagrams of
the types shown in the bottom part of Fig.~\ref{fig1}. Frankly speaking I've never
expected to discover a first order phase transition in liquid. But I
certainly hoped to find a maximum at the melting curve. I should mention
here the Gustave Tamman hypothesis on universal melting maxima, which is
certainly not true, and the wrong claim of a maximum on the Rb melting
curve, made by F.Bundy.
\begin{figure}[htb]
\includegraphics[width=80mm]{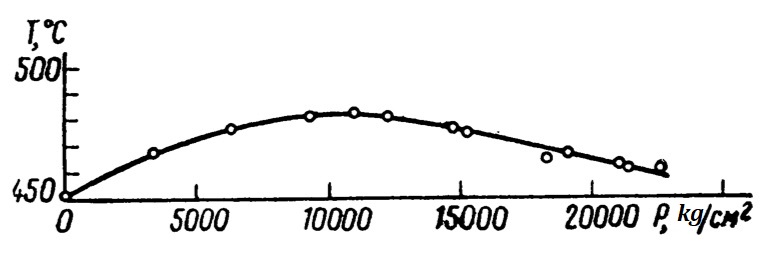}
\caption{\label{fig2} The melting curve of tellurium.The first observation of melting curve maxima~\cite{Tik}.}
\end{figure}
I am not going to describe here which way I followed to start studying the
phase diagram of tellurium. What is important that a maximum on the melting
curve of Te was discovered practically simultaneously in our paper~\cite{Tik}
(Fig.~\ref{fig2}) and in the paper of Kennedy's group~\cite{Ken}. In the same paper Kennedy
et al. announced a discovery of a double maximum at the melting curve of Cs.
In a while the phase diagram of Te was studied in more details~\cite{stish2} (Fig.~\ref{fig3}).
Later a rapid change of the slope of the Te melting curve was noticed at
pressures $\sim$5kb~\cite{stish3} (Fig.~\ref{fig4}) and I decided to look for a possible sharp
transformation in liquid Te. The easiest way was to measure the electrical
resistance of Te melt under pressure~\cite{stish4}. Though I should say it wasn't easy
at all, because of chemical reactivity of liquid Te. It corrodes almost
everything.
\begin{figure}[htb]
\includegraphics[width=80mm]{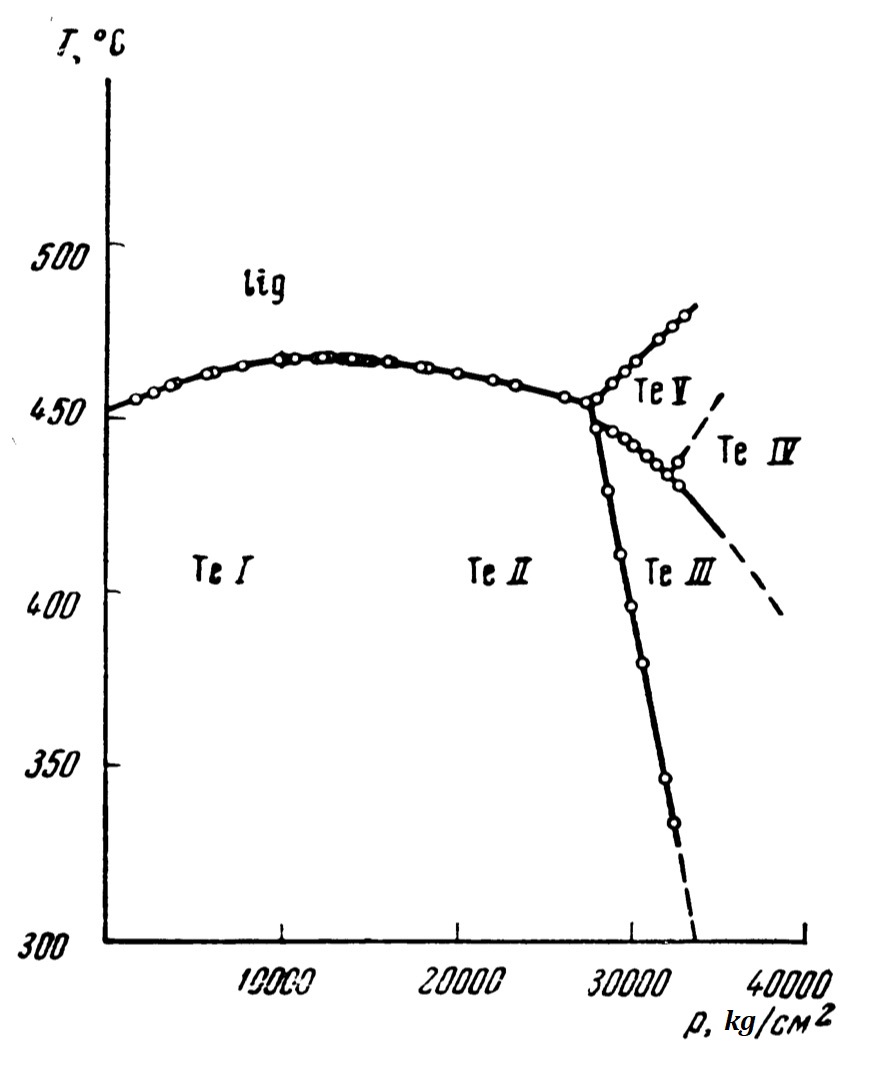}
\caption{\label{fig3} The phase diagram of tellurium~\cite{stish2}.}
\end{figure}
\begin{figure}[htb]
\includegraphics[width=80mm]{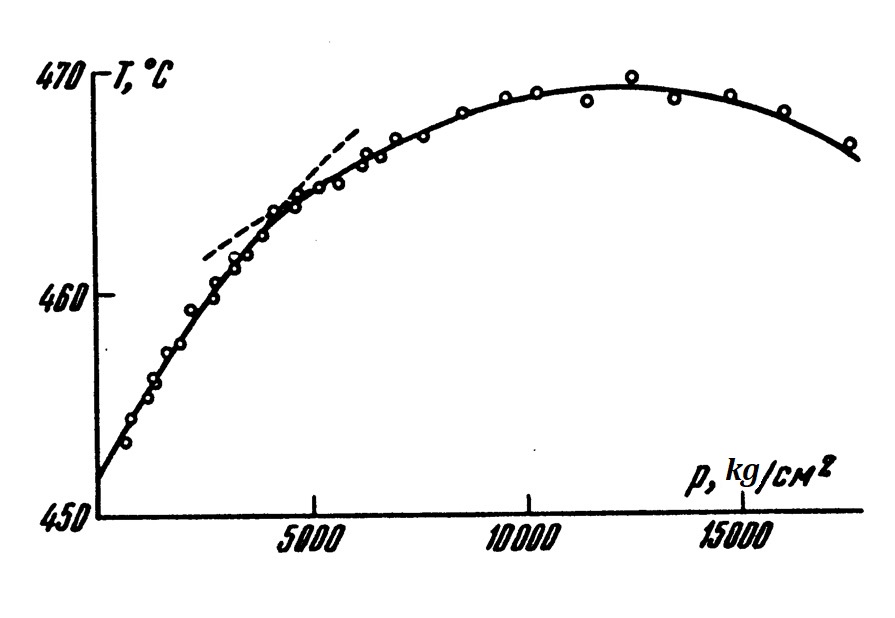}
\caption{\label{fig4} The rapid change of slope of melting curve of tellurium~\cite{stish3}.}
\end{figure}
The dependence of the electrical resistivity of liquid Te on pressure at 475
C is showed in Fig.~\ref{fig5}. This Figure is based on the data I obtained in 1966,
which never were published in this form before.
\begin{figure}[htb]
\includegraphics[width=80mm]{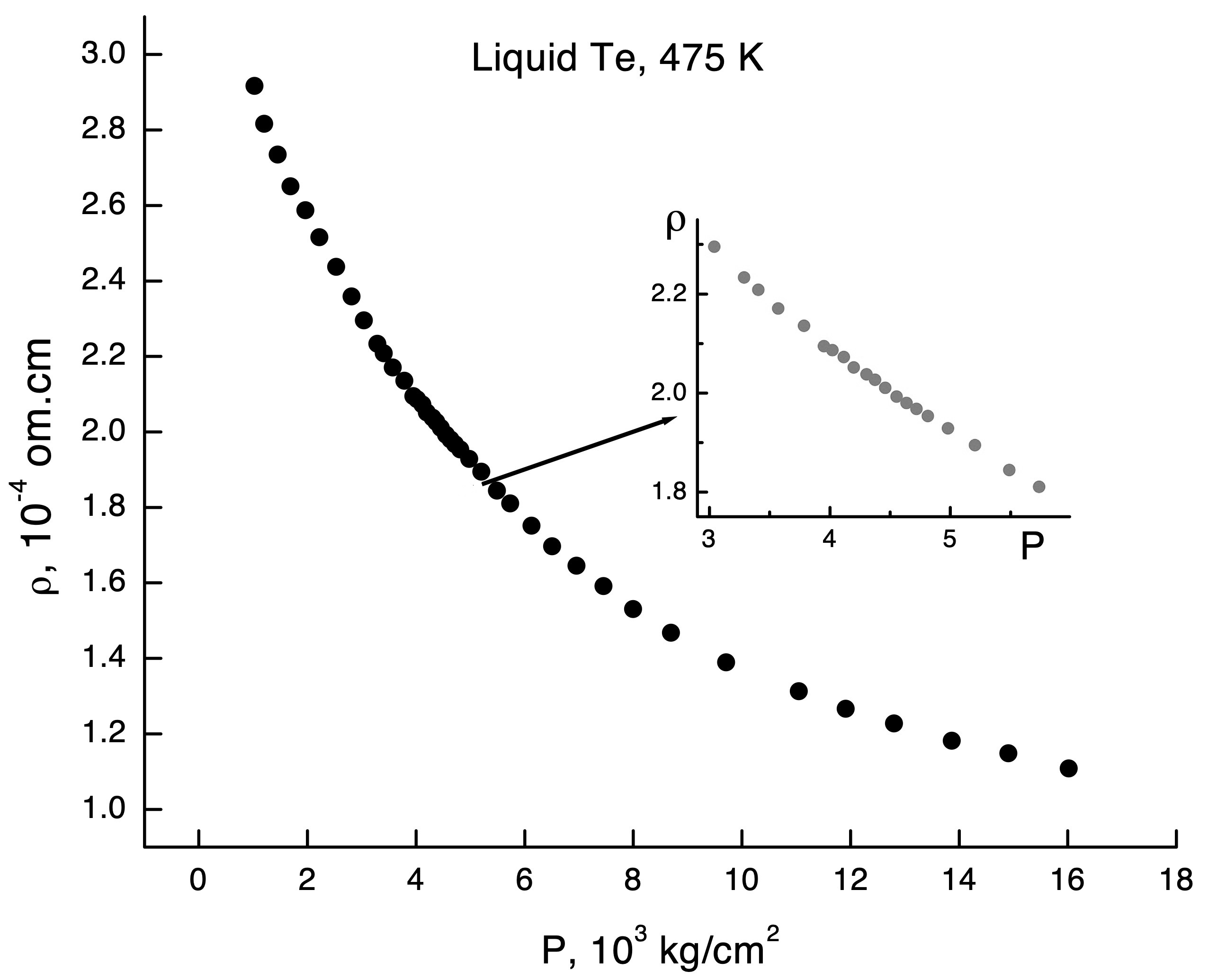}
\caption{\label{fig5} The resistivity of liquid Te at high pressure~\cite{stish4}.}
\end{figure}
As I remember it was a big disappointment for me to see just a smooth
variation of the resistivity with no indication for any localized change of
properties of liquid Te even in the region under suspicion. My attempts to
find evidences of abnormal behavior, manifesting phase transformations in
liquid Te under pressure, were not persuasive even for me~\cite{stish4}.
\begin{figure}[htb]
\includegraphics[width=80mm]{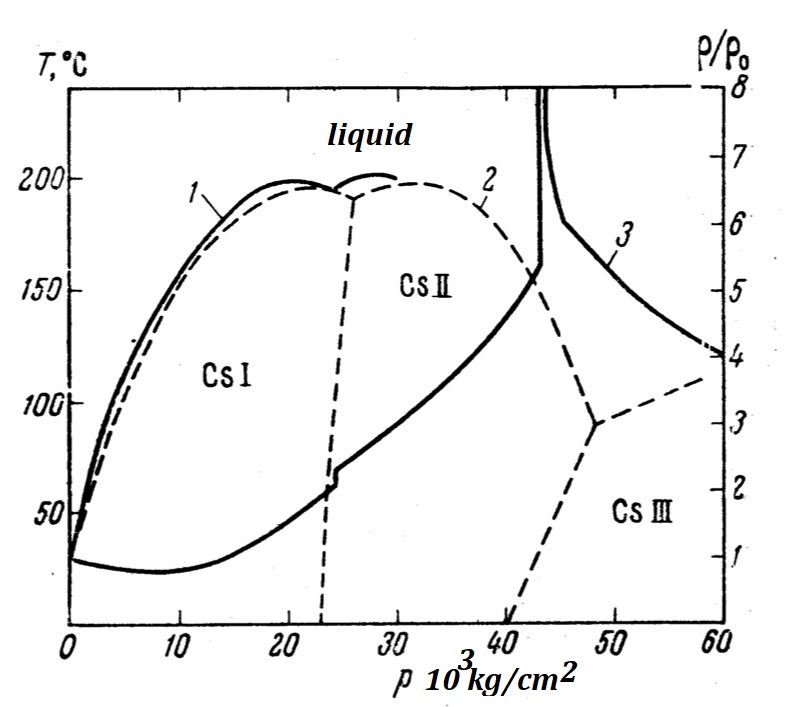}
\caption{\label{fig6} The phase diagram of cesium and the resistance curve of solid cesium, (1)\cite{Mak}, (2)\cite{Ken2}, (3)\cite{Hall}.}
\end{figure}
Next logical step at that time was to try to find out whether the melting
maxima of Cs are connected with some sort of localized transformations in
liquid Cs~Fig.~\ref{fig6}. The electronic nature of the phase transformations in solid Cs
gave some hope in this respect. Measuring of the electrical resistance of
liquid Cs was carried out~\cite{Mak}. Analogous measurements were published a
little bit earlier in~\cite{Jay}. Again as is seen in Fig.~\ref{fig7}, only smooth variation
of the resistivity of liquid Cs under pressure could be detected over
significant pressure range. On the other hand it's obvious that resistivity
of liquid and solid Cs displays quite anomalous features on compression,
which could be ascribed to the continuous change of the electronic structure.
\begin{figure}[htb]
\includegraphics[width=80mm]{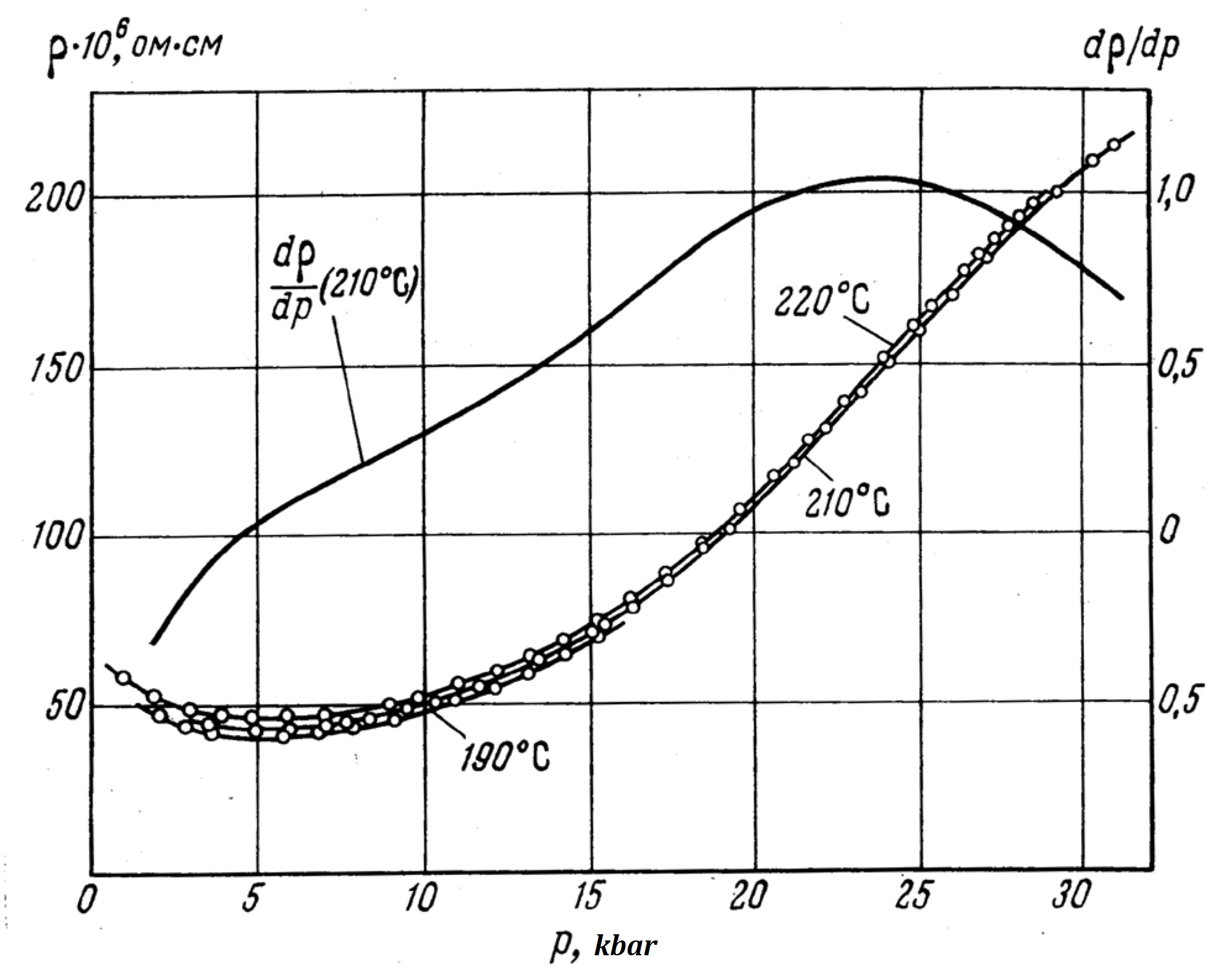}
\caption{\label{fig7} The resistivity of liquid cesium at high pressure after~\cite{Mak}.}
\end{figure}
A few years later we studied behavior of the thermodynamic quantities along
melting curves of Cs up to 22 kb using a dilatometrical technique~\cite{Mak2}
(Fig.~\ref{fig8}). Once again we could not find any localized anomaly, which might
indicate the phase transition in liquid, though there is no doubt that a
continuous transformation in liquid Cs occurred in the significant range of
pressure. The same conclusions were made upon completion of studying the
equation of state of liquid Cs~\cite{Mak3} (Fig.\ref{fig9}).
\begin{figure}[htb]
\includegraphics[width=80mm]{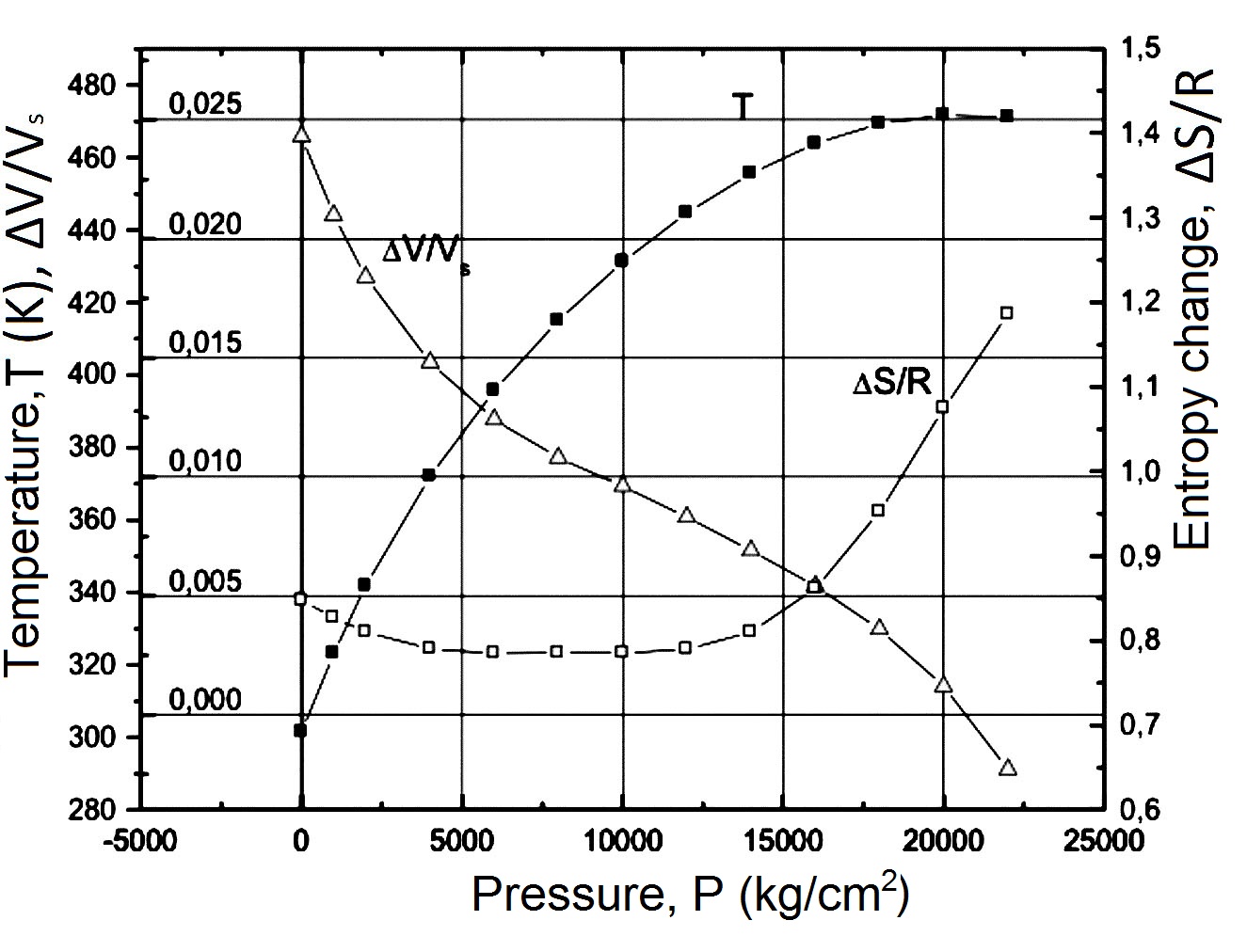}
\caption{\label{fig8} The melting properties of cesium~\cite{Mak2}.}
\end{figure}
\begin{figure}[htb]
\includegraphics[width=80mm]{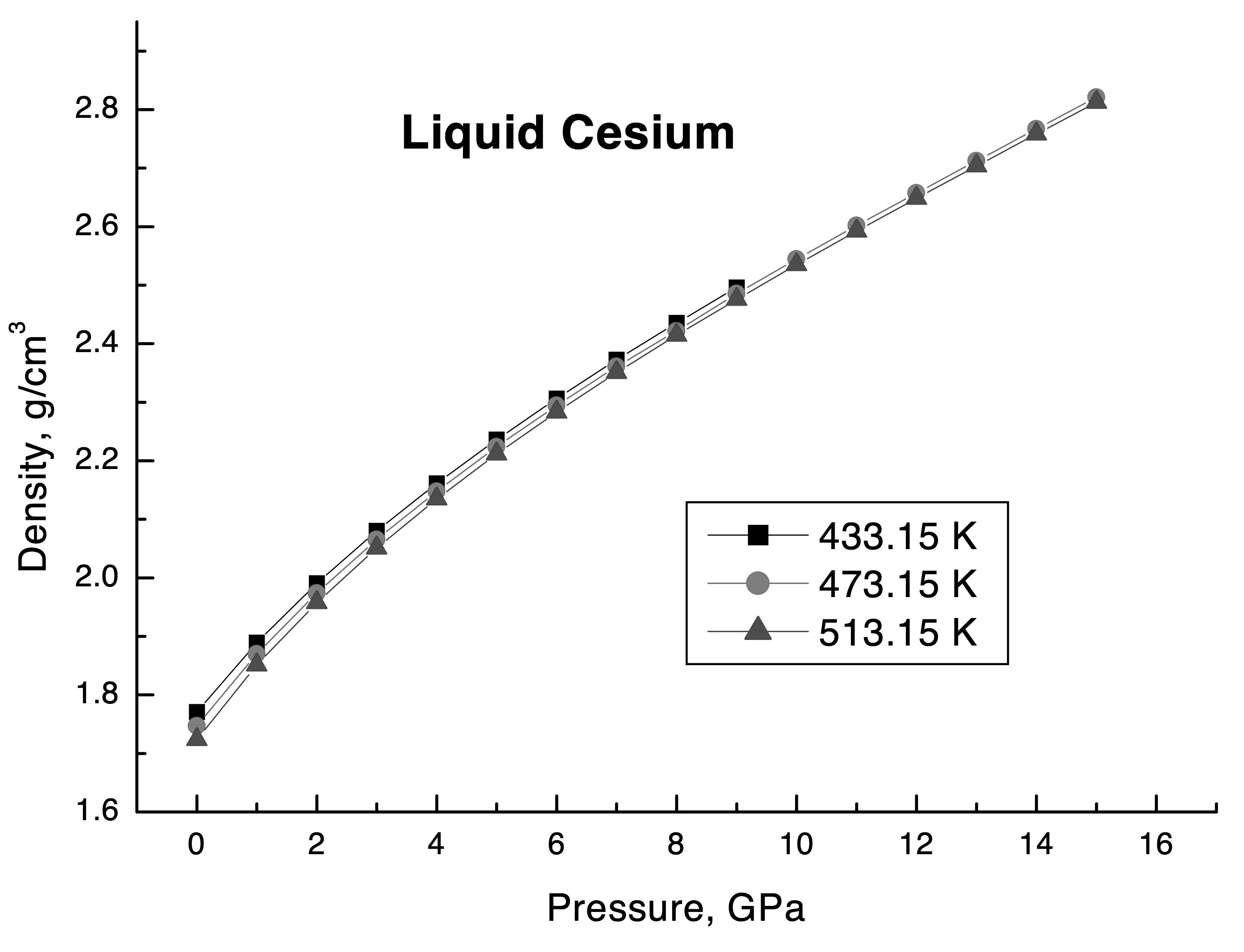}
\caption{\label{fig9} Density variation of liquid cesium at high pressure~\cite{Mak3}.}
\end{figure}
My involvement in the anomalous melting study was described in the review paper of 1968~\cite{stish5}.  
So I must say that my personal experience didn't favor the existence of
phase transitions in liquids. Of course it wasn't mean that they didn't
exist.
\begin{figure}[htb]
\includegraphics[width=80mm]{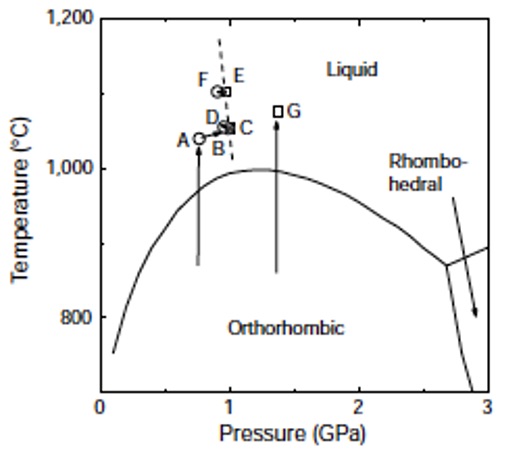}
\caption{\label{fig10} Location of the liquid-liquid phase transition on phase diagram of phosphorus.\cite{Kat6}}
\end{figure}

\begin{figure}[!h]
\includegraphics[width=80mm]{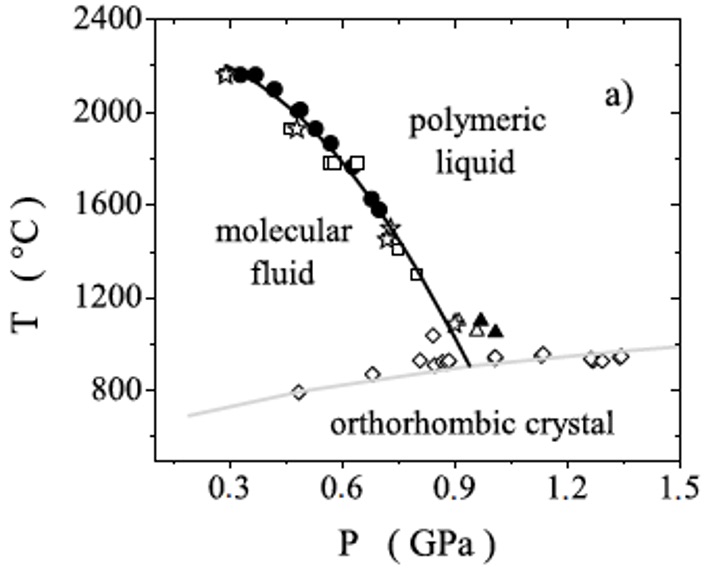}
\caption{\label{fig11} Partial phase diagram of phosphorus ~\cite{Mez}. 
Phase transition line between two liquid phase, which probably 
ends in the critical point,  is shown.}
\end{figure}

\begin{figure}[!h]
\includegraphics[width=80mm]{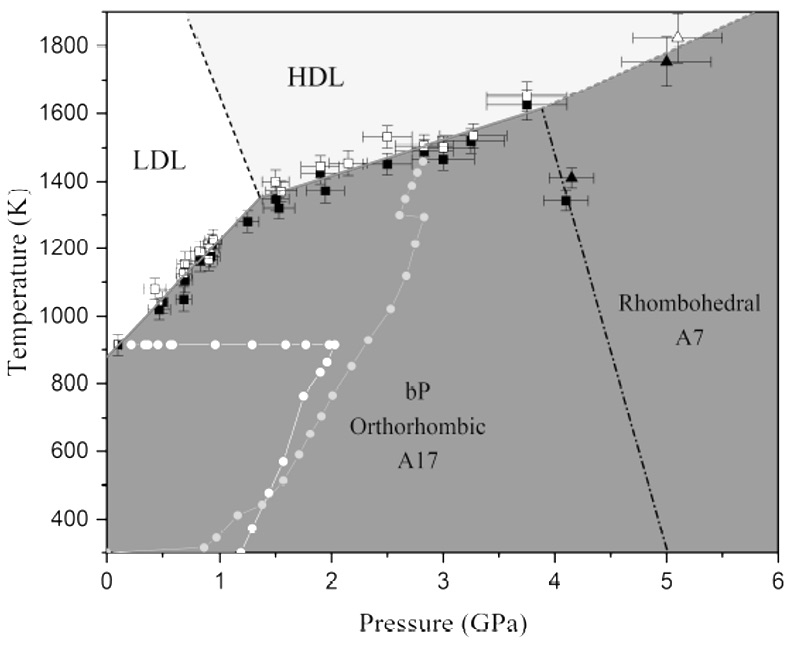}
\caption{\label{fig12} Melting line of black phosphurus after.~\cite{Mez2}}
\end{figure}

\begin{figure}[!h]
\includegraphics[width=80mm]{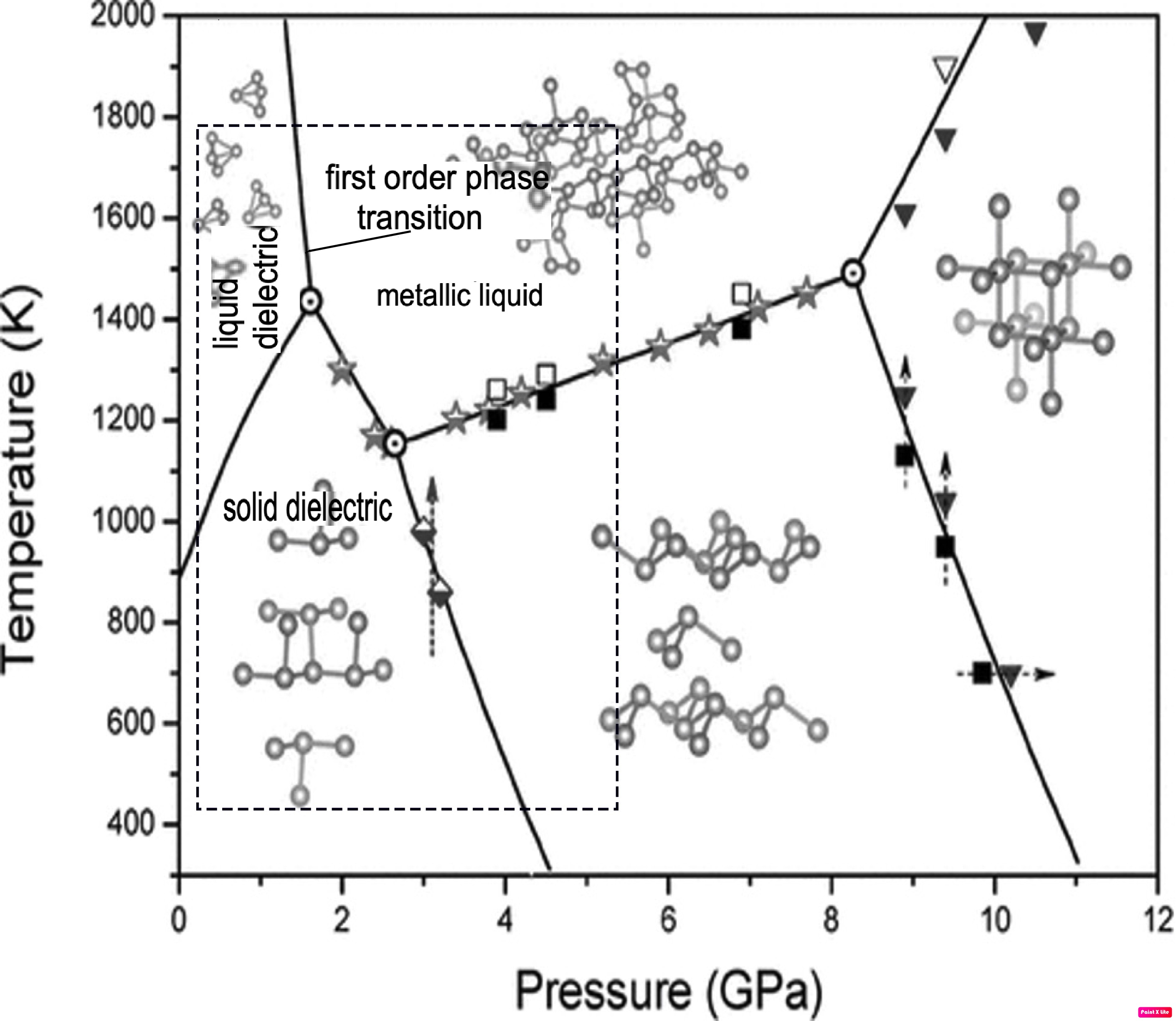}
\caption{\label{fig13} Phase diagram of phosphurus after.~\cite{Sol} It is amazing that the part of the diagram concluded 
in the rectangle is much simular to the diagram at the low right corner of Fig.1.   }
\end{figure}

Anyway in 1989  twenty years later since publication of my review ~\cite{stish5}  a paper, 
claiming a sharp phase transition in liquid selenium was published in JETP Letters (see translation in ~\cite{Bra})~\cite{note2}. 
Somewhat later the same team published a paper~\cite{Bra2}, describing their would be discovery 
a first order phase transition in liquid tellurium.  Note that in both case authors ~\cite{Bra,Bra2} employed so called thermobaro analysis
(TBA) to estimate the volume change making use the difference in the pressure dependence of the EMFs of two types of thermocouples.
This technique is absolutely not reliable in conditions of non hydrostatic pressure and strong temperature gradient.
In fact, the mentioned conclusions on phase transitions in liquid selenium and tellurium didn't find the solid confirmations when probing with the adequate tools
like EXAFS, X-ray diffraction, X-ray absorption~\cite{Kat,Kat2,Kat3,Kat4,Kat5}, carried out mainly by O.Shimomura group in Japan.    
However, there arose in 1999 some hope that the first order phase transitions in liquid might exist. This hope was based on paper~\cite{Ree} reporting a first order phase transition in liquid carbon between $sp$ and $sp_{3}$ states from the classical molecular dynamics study. Unfortunately, this hope vanished when the quantum   molecular dynamics was applied to the same problem. No phase transitions in liquid carbon was found~\cite{Wu}. Meanwhile the Japanese group didn't give up hope and 
continue their pursuit. Finally they finished up with the sensational discovery of the first order phase transition in liquid phosphorus at pressure about 1GPa~\cite{Kat6}. The transition occurred between  the molecular liquid of tetrahedral $P_{4}$ molecules and the polymeric liquid. Later on this transition was studied with more details in paper~\cite{Mez} (see Fig.11).

Since then quite a number of theoretical calculations based on first principals were performed, describing of the liquid-liquid transtion in phosphorus (see one of the last paper~\cite{Par} and corresponding references therein).
The melting curve of phosphorus at high pressures, which was studied some years ago~\cite{Ak}(see also Fig.10), was revised quite recently~\cite{Mez2, Sol} (Figs.12,13) with somewhat controversial results. It is interesting to note that the part of the diagram in Fig.13 is almost equivalent to the schematic phase diagram with a phase transition in liquid suggested in Ref.~\cite{stish}(see Fig.1). 

It  became clear after the exciting discovery in liquid phosphorus that the liquid-liquid phase transitions probably could be found most exclusively in substances with molecular structures. So the first order phase transition was recently discovered in liquid sulfur, corresponding transformation the $S_{8}$ cyclic molecules to a polymeric liquid~\cite{Mez3}. The transition in sulfur ended up in a critical point at high temperatures as it occurred in phosphorus. Note that the critical points on the liquid-liquid transition lines belong to the liquid-gas class since there are no symmetry limitations.
Then the obvious candidate for that kind phenomena is liquid hydrogen at the metallization transition (see for instance~\cite{Cep}). However, so far the experimental data are too controversial to make a positive conclusion.


\begin{thebibliography}{99}
\bibitem{note} It is curious that Fig.1 was almost completely reproduced in the much later paper~\cite{Makov}. 
Accidental coincidence was of cause possible but not highly probable.
\end{thebibliography}
\end{document}